\documentclass{emulateapj}

\begin{document}
\slugcomment{Accepted to ApJ}

\title{ Hypervelocity Stars III.  The Space Density and Ejection History of Main 
Sequence Stars from the Galactic Center}

\author{Warren R.\ Brown,
	Margaret J.\ Geller,
	Scott J.\ Kenyon,
	Michael J.\ Kurtz}

\affil{Smithsonian Astrophysical Observatory, 60 Garden St, Cambridge, MA 02138}
\email{wbrown@cfa.harvard.edu}

\and	\author{Benjamin C.\ Bromley}
\affil{Department of Physics, University of Utah, 115 S 1400 E, Salt Lake City, UT 84112}

\shorttitle{ Hypervelocity Stars III. }
\shortauthors{ Brown et al. }

\begin{abstract}

	We report the discovery of 3 new unbound hypervelocity stars (HVSs), stars
traveling with such extreme velocities that dynamical ejection from a massive black
hole (MBH) is their only suggested origin.  We also detect a population of possibly
bound HVSs.  The significant asymmetry we observe in the velocity distribution -- we
find 26 stars with $v_{\rm rf}>275$ km s$^{-1}$ and 1 star with $v_{\rm rf}<-275$ km
s$^{-1}$ -- shows that the HVSs must be short-lived, probably 3 - 4 M$_{\sun}$ main
sequence stars.  Any population of hypervelocity post-main sequence stars should
contain stars falling back onto the Galaxy, contrary to the observations.
	The spatial distribution of HVSs also supports the main sequence
interpretation:  longer-lived 3 M$_{\sun}$ HVSs fill our survey volume;
shorter-lived 4 M$_{\sun}$ HVSs are missing at faint magnitudes.
	We infer that there are $96\pm10$ HVSs of mass 3 - 4 M$_{\sun}$ within
$R<100$ kpc, possibly enough HVSs to constrain ejection mechanisms and potential
models.  Depending on the mass function of HVSs, we predict that SEGUE may find up
to 5 - 15 new HVSs.  The travel times of our HVSs favor a continuous ejection
process, although a $\sim$120 Myr-old burst of HVSs is also allowed.

\end{abstract}


\section{INTRODUCTION}

	In 2005 we discovered the first HVS: a 3 M$_{\sun}$ main sequence star
traveling with a Galactic rest frame velocity of at least $+709\pm12$ km s$^{-1}$,
many times the escape velocity of the Galaxy at its heliocentric distance of 110 kpc
\citep{brown05, fuentes06}.  This star cannot be explained by normal stellar
interactions: the maximum ejection velocity from binary disruption mechanisms
\citep{blaauw61, poveda67} is limited to $\sim$300 km s$^{-1}$ for few M$_{\sun}$
stars \citep{leonard91, leonard93, tauris98, portegies00, davies02, gualandris05}.
Thus the origin of the HVS must be tied to a more massive and compact object.

	\citet{hills88} first predicted HVSs as an inevitable consequence of
three-body interactions with a massive black hole (MBH).  There is overwhelming
evidence for a $4\times10^6$ M$_{\sun}$ MBH at the center of our Galaxy
\citep{schodel03, ghez05}.  Thus HVSs are probably stars ejected from the Galaxy by
the MBH in the Galactic center.

	Further HVS discoveries have followed the original discovery.  
\citet{hirsch05} reported a helium-rich subluminous O star leaving the Galaxy with a
rest-frame velocity of at least $+717$ km s$^{-1}$.  \citet{edelmann05} reported an
8 M$_{\sun}$ main sequence star with a Galactic rest frame velocity of at least
$+548$ km s$^{-1}$, possibly ejected from the Large Magellanic Cloud.  
\citet{brown06, brown06b, brown07a} designed a targeted HVS survey using the 6.5m
MMT and Whipple 1.5m telescopes to measure the radial velocities of distant B-type
stars.  This strategy has worked remarkably well, yielding 7 HVSs and evidence for a
bound population of stars ejected by the same mechanism.  Here we report the 3
newest HVS discoveries, and discuss additional evidence for the ``bound'' HVSs.

	The existence of HVSs has inspired broad theoretical interest, and many
testable predictions have emerged.  It is clear that HVSs can be ejected by
different mechanisms: binary star encounters with a single MBH \citep{hills88, yu03,
bromley06} or possibly with an intermediate mass black hole \citep[IMBH,][]{
gualandris07}, single star encounters with a binary MBH \citep{yu03, sesana06,
merritt06, sesana07} or an in-spiraling IMBH-MBH \citep{gualandris05, levin06,
baumgardt06}, and single star encounters with the cluster of stellar mass black
holes around the MBH \citep{oleary07}.  The dominant ejection mechanism remains
unclear.  Ejection rates depend on the number of stars scattered onto orbits that
intersect the MBH's ``loss cone'' \citep{perets07, perets07b}.  Interestingly, the
different ejection mechanisms result in different distributions of HVS ejection
velocities and ejection rates \citep[e.g.][]{yu03, sesana07b}.

	HVSs probe a variety of characteristics of the Galaxy.  The density of HVSs
and their distribution of velocities tell us about the MBH's environment.  The
stellar types of HVSs tell us about the types of stars orbiting near the MBH
\citep{brown06, demarque07, kollmeier07, lu07}.  The trajectories of HVSs provide
unique probes of the shape and orientation of the Galaxy's dark matter halo
\citep{gnedin05, yu07}.  A large sample of HVSs will be a new and powerful tool to 
investigate these issues.

	Our HVS radial velocity survey is now 96.5\% complete for faint B-type stars
over 7300 deg$^2$ of the Sloan Digital Sky Survey (SDSS) Data Release 5
\citep[DR5,][]{adelman07}.  Our survey provides strong evidence for a class of HVSs
on bound orbits \citep{brown07a} matching expectations from theoretical models
\citep{bromley06, sesana07b}.  The distribution of HVSs on the sky appears
marginally anisotropic, as expected for a magnitude-limited survey of HVSs ejected
from the Galactic center \citep{brown07a}.

	Here we address the HVS's stellar type, their space density, and the history
of HVS ejections from the Galactic center.
	In \S 2 we discuss the final target selection and spectroscopic
identifications.  In \S 3 we describe three new HVSs. In \S 4 we show that the
observed HVSs must be short-lived objects, probably main sequence stars.  In \S 5 we
calculate the space density of HVSs, and predict HVSs discovery rates for some
future surveys.  In \S 6 we calculate HVS travel times, and discuss their ejection
history.  We conclude in \S 7.  The new observations are in Appendix A.

\section{DATA}

\subsection{Target Selection}

	\citet[hereafter Paper I]{brown06b} describe the target selection for our
survey of faint $17<g'_0<19.5$ HVS candidates.  Briefly, we use SDSS photometry to
select HVS candidates with the colors of late B-type stars.  B-type stars have
lifetimes consistent with travel times from the Galactic center but are not a
normally expected Galactic halo population.  Our color selection introduces no
kinematic bias and allows for stars at any velocity.

\begin{figure}          
 \includegraphics[width=3.25in]{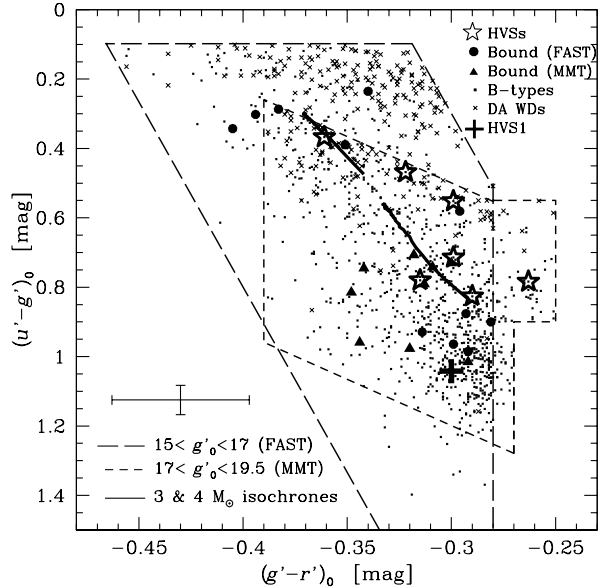}
 \caption{ \label{fig:ugr}
        Color-color diagram showing the spectroscopic identifications of the 1005  
late-B spectral type stars ({\it squares}) and the 244 white dwarfs ({\it x's}) in
our survey.  Target selection regions are shown for the MMT ({\it short dashed
line}) and FAST ({\it long dashed line}) samples.  If the HVSs ({\it stars}) are  
main sequence stars, then \citet{girardi04} stellar evolution tracks suggest they
are 3 M$_{\sun}$ ({\it lower solid line}) or 4 M$_{\sun}$ stars ({\it upper solid 
line}).  The average color uncertainties are illustrated by the errorbar on the
lower left.  Also marked are the 19 possibly bound HVSs in the MMT sample ({\it
triangles}) and FAST sample ({\it circles}), and HVS1 ({\it plus sign}).}
 \end{figure}

	In this paper we make two changes to the Paper I target selection:  1) we
select targets from the larger SDSS DR5 catalog, and 2) we include stars in a
supplementary color-color region defined by $0.55<(u'-g')_0<0.9$ and
$-0.28<(g'-r')_0<-0.25$ (see Figure \ref{fig:ugr}).
	Our full target list contains 621 HVS candidates over 7300 deg$^2$ of sky.  
We have observed 575 of these targets; we are thus 93\% complete for the faint HVS
candidates.  The remaining 46 targets are confined to the region $13\fh25 < {\rm RA}
< 14\fh5$.

	\citet[hereafter Paper II]{brown07a} describe the target selection for our
complementary survey of bright $15<g'_0<17$ HVS candidates.  We now select targets
from the SDSS DR5 catalog.  We identify a total of 892 objects with B-type colors;
691 satisfy our Galactic longitude and latitude cut.  We have observed all 691
objects and are thus 100\% complete for bright HVS candidates.  These objects are
spread over the identical 7300 deg$^2$ of sky as the faint HVS candidates.

\subsection{Spectroscopic Observations}

	New observations of faint HVS candidates were obtained at the 6.5m MMT
telescope with the MMT Blue Channel spectrograph on the nights of 2006 December
21-26, 2007 March 18-20, and 2007 May 14-18.  We operated the spectrograph with the
832 line mm$^{-1}$ grating in second order and a 1.25$\arcsec$ slit.  These settings
provide wavelength coverage 3650 \AA\ to 4500 \AA\ and a spectral resolution of 1.2
\AA.  Exposure times ranged from 5 to 30 minutes and yielded signal-to-noise
$S/N=15$ in the continuum at 4000 \AA.  We obtained comparison lamp exposures after
every exposure.

	We also obtained new observations of bright HVS candidates at the Whipple
Observatory 1.5m Tillinghast telescope with the FAST spectrograph
\citep{fabricant98} over the course of 11 nights spread between 2006 September and
2007 May.  We operated the spectrograph with a 600 line mm$^{-1}$ grating and a
2$\arcsec$ slit.  These settings provide wavelength coverage 3500 \AA\ to 5400 \AA\
and a spectral resolution of 2.3 \AA.  Like the MMT observations, exposure times
ranged from 5 to 30 minutes and yield $S/N=15$ in the continuum at 4000 \AA.  We
obtained comparison lamp exposures after every exposure.

	We extracted the spectra using IRAF\footnote{IRAF is distributed by the
National Optical Astronomy Observatories, which are operated by the Association of
Universities for Research in Astronomy, Inc., under cooperative agreement with the
National Science Foundation.}
	in the standard way and measured radial velocities using the
cross-correlation package RVSAO \citep{kurtz98}.  \citet{brown03} describe the
cross-correlation templates we use.  The radial velocity accuracy is $\pm11$ km
s$^{-1}$ for the B-type stars.

\subsection{Spectroscopic Identifications}

	Figure \ref{fig:ugr} plots the colors and spectroscopic identifications of 
the 1,266 HVS candidates in our survey.  We find that 1005 (79.4\%) are stars of B 
spectral type and 244 (19.3\%) are DA white dwarfs.
	Curiously, the white dwarfs in our survey have very unusual colors for DA
white dwarfs; these colors imply very low surface gravity.  Indeed, a
detailed analysis of the spectra reveal that one of the white dwarfs has a mass of
0.17 M$_{\sun}$, the lowest mass white dwarf ever found \citep{kilic07, kilic07b}.  
Other remarkable objects include B supergiants in the Leo A dwarf galaxy
\citep{brown07c} and one compact, extremely metal poor galaxy with properties
similar to the host galaxies of nearby gamma ray bursts \citep{kewley07, brown07d}.

	Our new observations include a B supergiant in the Sextans B dwarf galaxy.  
The star SDSS J095951.18+052124.52 is located within the galaxy's $3\arcmin$
Holmberg radius \citep{mateo98} and has a $+295\pm11$ km s$^{-1}$ heliocentric
radial velocity, consistent with the $+295\pm10$ km s$^{-1}$ systemic velocity of
Sextans B \citep{fisher75}.  The observed spectrum is that of a luminosity class I B
supergiant.  The $(m-M)=25.6$ distance modulus of Sextans B \citep{piotto94,
mendez02} implies the star has $M_V=-6.25$, also consistent with a B I supergiant.  
We find no other B supergiants in our survey, suggesting that, to the depth of our
survey, the census of star-forming dwarf galaxies in the Local Group may be complete
over the region of sky covered by SDSS DR5.

	In Figure \ref{fig:ugr} we also plot \citet{girardi02, girardi04} main
sequence track for a solar abundance 3 M$_{\sun}$ star (the lower solid line) and a
4 M$_{\sun}$ star (the upper solid line).  If main sequence stars fall in our
survey, our color selection targets primarily 3 - 4 M$_{\sun}$ stars.
	All the HVSs overlap the main sequence tracks within their 1-$\sigma$
uncertainties ($\sigma_{u'-g'}=\pm0.042$, $\sigma_{g'-r'}=\pm0.033$). 

\subsection{Radial Velocity Distribution and Possibly Bound HVSs}

\begin{figure}          
 \includegraphics[width=3.25in]{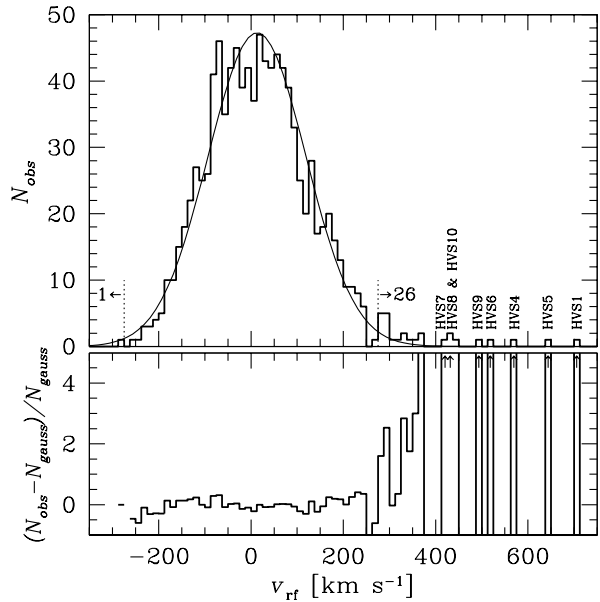}
 \caption{ \label{fig:hist}
        Galactic rest-frame velocity histogram for all 1005 B-type stars in our
sample ({\it upper panel}).  The best-fit Gaussian ({\it thin line}) has dispersion
$106\pm4$ km s$^{-1}$.  The lower panel plots the residuals of the observations from
the best-fit Gaussian, normalized by the value of the Gaussian.  The strong    
asymmetry of stars with $|v_{rf}|>275$ km s$^{-1}$ is significant at the
$\sim$$5\sigma$ level, and shows that the bound stars with $+300$ km s$^{-1}$ are
{\it short-lived}; we observe only 1 star falling back onto the Galaxy near $-300$
km s$^{-1}$. }
 \end{figure}

	Figure \ref{fig:hist} plots the distribution of line-of-sight velocities,
corrected to the Galactic rest-frame $v_{\rm rf}$, for our entire sample of 1005
B-type stars.  We calculate minimum Galactic rest-frame velocities by removing the
radial component of the Sun's 220 km s$^{-1}$ orbital motion and the radial
component of the Sun's motion with respect to the Local Standard of Rest from the
observed radial velocities (see Paper I).

	The tail of positive velocity outliers in Figure \ref{fig:hist} is truly
striking:  we observe 26 stars with $v_{\rm rf}>275$ km s$^{-1}$ and only 1 star
with $v_{\rm rf}<-275$ km s$^{-1}$.
	We estimate the significance of the asymmetric velocity distribution by
fitting a Gaussian distribution to the data.  We iteratively clip 3-$\sigma$
outliers and find a $+13\pm3$ km s$^{-1}$ mean and a $106\pm4$ km s$^{-1}$
dispersion.  The lower panel of Figure \ref{fig:hist} plots the residuals of the
observations from this Gaussian distribution, normalized by the value of the
Gaussian.  Stars with velocities $|v_{\rm rf}|<275$ km s$^{-1}$ show low-significance
deviations from a Gaussian distribution and are likely a normal halo population
(Paper I).  Integrating the tail of the Gaussian suggests we should expect 7 stars
with $v_{\rm rf}>275$ km s$^{-1}$.  Yet we observe 26 such objects, including the
unbound HVSs.
	There is less than a $10^{-7}$ probability of drawing 26 stars with
$v_{\rm rf}>275$ km s$^{-1}$ from a Gaussian distribution with the observed parameters.  
Thus the observed asymmetry appears significant at the 5-$\sigma$ level.

	We now ask whether the non-zero mean of the distribution is consistent with
models of HVS ejections.  If the 19 objects in excess of the Gaussian distribution
$v_{\rm rf}>275$ km s$^{-1}$ are HVSs, the \citet{bromley06} ejection model for 3
M$_{\sun}$ stars suggests that an additional 11 HVSs have $v_{\rm rf}<275$ km s$^{-1}$
in our survey.  We randomly draw 30 objects from the \citet{bromley06} HVS ejection
model and add them to 975 objects drawn from a Gaussian distribution with dispersion
106 km s$^{-1}$ and zero mean.  The expected 30 HVSs shift the mean of the
distribution by $+2\pm4$ km s$^{-1}$, consistent with but not entirely sufficient to
explain the observed shift in the mean.

	In Paper II, we argue that the significant excess of stars with velocities
around $v_{\rm rf}\sim +300$ km s$^{-1}$ demonstrates a population of HVSs ejected
onto bound trajectories.  HVS ejection mechanisms naturally produce broad spectrum
of ejection velocities that include bound orbits \citep{bromley06}.  We observe 19
possibly bound HVSs with $275<v_{\rm rf}<400$ km s$^{-1}$ in our completed survey.  
The colors of the bound HVSs suggest they are a mixed population: some of the bound
HVSs are significantly redder or bluer than the main sequence tracks (see Figure
\ref{fig:ugr}).  However, it is possible to pick a sample of $\sim$12 bound HVSs,
the number in excess of the Gaussian distribution, that overlap the color
distribution of the unbound HVSs.  Follow-up spectroscopy is therefore required to
constrain the stellar type of the possibly bound HVSs.  Because we do not know which
of the possibly bound HVSs are true HVSs, we focus our attention on the well-defined
sample of unbound HVSs in this paper.

\begin{deluxetable*}{lcccccccccl}           
\tablewidth{0pt}
\tablecaption{HYPERVELOCITY STARS\label{tab:hvs}}
\tablecolumns{11}
\tablehead{
  \colhead{ID} & \colhead{Type} & \colhead{$M_V$} & \colhead{$V$} &
  \colhead{$R_{GC}$} & \colhead{$l$} & \colhead{$b$} & \colhead{$v_{\sun}$} &
  \colhead{$v_{\rm rf}$} & \colhead{$t_{GC}$} & \colhead{Catalog} \\
  \colhead{} & \colhead{} & \colhead{{\small mag}} & \colhead{{\small mag}} &
  \colhead{{\small kpc}} & \colhead{{\small deg}} & \colhead{{\small deg}} & 
  \colhead{{\small km s$^{-1}$}} & \colhead{{\small km s$^{-1}$}} &
  \colhead{{\small Myr}} & \colhead{}
}
	\startdata
hvs1  &  B  & -0.3 & 19.84 & 111 & 227.33 & +31.33 & 840 & 696 & 145 & SDSS J090744.99+024506.9$^1$ \\
hvs2  & sdO & +2.6 & 19.05 &  26 & 175.99 & +47.05 & 708 & 717 & 32  & US 708$^2$                   \\
hvs3  &  B  & -2.7 & 16.20 &  62 & 263.04 & -40.91 & 723 & 548 & 100 & HE 0437-5439$^3$             \\
hvs4  &  B  & -0.9 & 18.50 &  82 & 194.76 & +42.56 & 611 & 567 & 125 & SDSS J091301.01+305119.8$^4$ \\
hvs5  &  B  & -0.3 & 17.70 &  45 & 146.23 & +38.70 & 551 & 647 & 60  & SDSS J091759.48+672238.3$^4$ \\
hvs6  &  B  & -0.3 & 19.11 &  78 & 243.12 & +59.56 & 626 & 528 & 130 & SDSS J110557.45+093439.5$^5$ \\
hvs7  &  B  & -0.9 & 17.80 &  56 & 263.83 & +57.95 & 534 & 421 & 110 & SDSS J113312.12+010824.9$^5$ \\
hvs8  &  B  & -0.3 & 18.09 &  53 & 211.70 & +46.33 & 511 & 429 & 100 & SDSS J094214.04+200322.1     \\
hvs9  &  B  & -0.3 & 18.76 &  68 & 244.63 & +44.38 & 628 & 485 & 120 & SDSS J102137.08$-$005234.8   \\
hvs10 &  B  & -0.3 & 19.36 &  87 & 249.93 & +75.72 & 478 & 432 & 165 & SDSS J120337.85+180250.4     \\
	\enddata
\tablerefs{ (1) \citet{brown05}; (2) \citet{hirsch05}; (3) \citet{edelmann05};
(4) \citet{brown06}; (5) \citet{brown06b} }
 \end{deluxetable*}

\section{HYPERVELOCITY STARS}

\subsection{ Three New HVSs }

	We discover 3 HVSs in our new observations.  The objects are SDSS
J094214.04+200322.07 (HVS8), SDSS J102137.08-005234.77 (HVS9), and SDSS
J120337.86+180250.35 (HVS10).  HVS8 has a B9 spectral type, a $+512\pm10$ km
s$^{-1}$ heliocentric radial velocity, and a minimum velocity of $+430$ km s$^{-1}$
in the Galactic rest frame.  A solar metallicity 3 M$_{\sun}$ main sequence star has
$M_V\simeq-0.3$ \citep{schaller92}.  This luminosity places HVS8 at a Galactocentric
distance $R=53$ kpc.  
	The mass of the Galaxy within 50 kpc is $5.5\times10^{11}$ M$_{\sun}$
\citep{wilkinson99, sakamoto03}, which implies that the the escape velocity at 50
kpc is $\sim$300 km s$^{-1}$.
	We conclude that HVS8 is very likely unbound to the Galaxy.

	HVS9 has a B9 spectral type, a $+632\pm11$ km s$^{-1}$ heliocentric radial
velocity, and a minimum velocity of $+489$ km s$^{-1}$ in the Galactic rest frame.  
	Unlike the other HVSs in our survey, which are located 10\arcdeg\ -
20\arcdeg\ away on the sky from the nearest Local Group galaxy, HVS9 is 2\fdg3 from
the Sextans dwarf galaxy.  Any physical association with Sextans is very unlikely,
however. Sextans is $1320\pm40$ kpc distant \citep{dolphin03} and has a heliocentric
velocity of $224\pm2$ km s$^{-1}$ \citep{young00}.  Thus HVS9 is moving toward the
dwarf galaxy with a relative velocity of 408 km s$^{-1}$.
	Interestingly, HVS9 also has the reddest $(g'-r')$ color of the HVSs; it is
arguably the best blue horizontal branch (BHB) candidate among our HVSs.  The
equivalent width of its Ca {\sc ii} K line suggests low metallicity and the
\citet{clewley04} line-shape technique suggests low surface gravity.  However, both
measurements are notoriously uncertain at the hot effective temperature of HVS9,
$(B-V)\simeq-0.01$.  Whether HVS9 is a BHB star located at $R=35$ kpc or a 3
M$_{\sun}$ main sequence star at $R=68$ kpc, its large velocity means it is
certainly unbound.

	HVS10 has a B9 spectral type, a $+478\pm10$ km s$^{-1}$ heliocentric radial
velocity, and a minimum velocity of $+432$ km s$^{-1}$ in the Galactic rest frame.  
HVS10 is the faintest HVS discovered by this survey, $g'=19.295\pm0.024$, and is
thus the most distant HVS known except for HVS1.  Assuming it is a 3 M$_{\sun}$ main
sequence star, HVS10 is $\sim$84 kpc above the Galactic plane at its latitude
$b=+76\arcdeg$.

	Table \ref{tab:hvs} summarizes the HVS properties.  Columns include HVS
number, stellar type, inferred absolute magnitude $M_V$, apparent $V$ magnitude
derived from SDSS photometry, Galactocentric distance $R$, Galactic coordinates
$(l,b)$, heliocentric radial velocity $v_{\sun}$, minimum Galactic rest-frame
velocity $v_{\rm rf}$ (not a full space velocity), travel time estimate from the
Galactic Center $t_{GC}$, and catalog identification.  Velocities have changed
slightly from previous work because we have obtained multiple observations of all
the HVSs; here we report the weighted average of the velocity measurements.  We
include HVS1, HVS2, and HVS3 in Table \ref{tab:hvs} but exclude them from our
analyses; the first three HVSs do not fall within the color/magnitude bounds of this
targeted survey.

	We do not report errors in Table \ref{tab:hvs} because formal uncertainties
are misleadingly small compared to the (unknown) systematic errors.  For example,
heliocentric radial velocities are accurate to $\pm$11 km s$^{-1}$, but we have no
constraint on the proper motion component of the rest frame velocity $v_{\rm rf}$.  
The luminosity estimates described below are precise at the 10\% level for main
sequence stars, however the luminosity estimates could be over-estimated by an order
of magnitude for post-main sequence stars.

\subsection{ Constraints on HVS Binaries }

	Approximately 25\% of late B-type stars in the solar neighborhood are in
binaries with periods $<100$ days and with mass ratios greater than $m_2/m_1 > 0.1$
\citep{wolff78}.  Thus, in principle, $2\pm1$ of the 8 late B-type HVSs (we include
HVS1) could be spectroscopic binaries.  Although HVSs are ejected as single stars by
standard HVS mechanisms, \citet{lu07} argue that the MBH binary mechanism can
eject a compact HVS binary.  Detection of a single HVS binary might thus provide
compelling evidence for a binary MBH in the Galactic center \citep{lu07}.

	We have two spectroscopic observations of each of the 7 HVSs in this survey,
typically obtained a few months apart.  In all cases, the radial velocities are
identical within the accuracy of the measurements, $\pm11$ km s$^{-1}$.  Thus it
appears unlikely that the HVSs are compact binaries.  We test this claim by
comparing the observations with velocities drawn $10^4$ times from a typical 50 km
s$^{-1}$ semi-amplitude binary with random orbital phase and inclination.  A
Kolmogorov-Smirnov (KS) test finds a 0.04 likelihood of drawing the observations
from a binary system with 50 km s$^{-1}$ semi-amplitude.  Additional observations of
higher accuracy are necessary to rule out radial velocity variations completely.

	We have also obtained three observations of HVS1 spread over 3.5 years.  
The heliocentric radial velocities are $853\pm12$, $816\pm14$, and $832\pm13$ km
s$^{-1}$.  Thus HVS1 may have velocity variations.  However, HVS1 is a slowly
pulsating B variable \citep{fuentes06}.  Slowly pulsating B variables exhibit radial
velocity amplitudes of $\sim$20 km s$^{-1}$ \citep{aerts99, mathias01}, consistent
with the observations.  Although our current observations are not conclusive, it is
unlikely that any of the known HVSs are compact binary systems.

\begin{figure}		
 \includegraphics[width=3.25in]{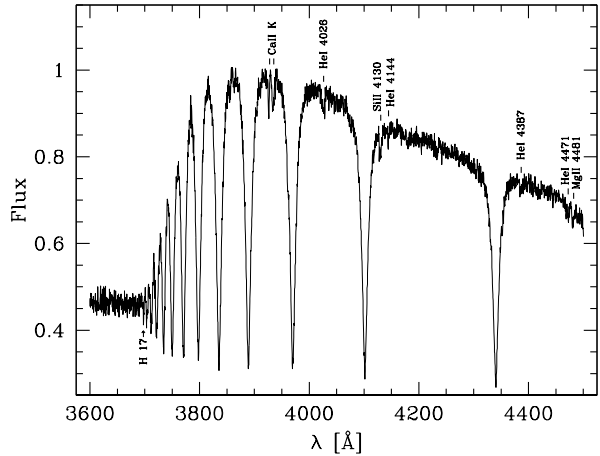}
 \caption{ \label{fig:spectrum}
	Summed HVS spectrum, created from the weighted average of our observations
of HVS1 and HVS4 - HVS10, shifted to rest frame.  The spectral type is that of a B8
- B9 star.  The wavelength difference between the pair of Ca {\sc ii} K lines, one
from the HVSs and one from the local interstellar medium that appears in the 
spectra of HVS8 and HVS9, visibly indicates the large space motion of the HVSs, 
$\Delta \lambda / \lambda \sim$ 550 km s$^{-1}$.}
 \end{figure}

\subsection{ HVS Template Spectrum }

	Summing our HVS spectra allows construction of a high $S/N$ template of the
HVSs.  Figure \ref{fig:spectrum} displays the weighted average of our observations
of HVS1 and HVS4 - HVS10, shifted to the rest frame.  The total integration time is
equivalent to 4 hours on the 6.5m MMT telescope.  The strongest spectral features in
Figure \ref{fig:spectrum} are the hydrogen Balmer lines, visible from H$\gamma$ to H17.

	It is interesting to discuss spectral classification in light of the weak
spectral lines visible in Figure \ref{fig:spectrum}.  The strengths of Mg {\sc ii}
4481 and the Si {\sc ii} 4130 blend relative to He {\sc i} 4471 and He {\sc i} 4144,
respectively, indicate that the summed HVS spectrum must be later than a B7 spectral
type.  Notably, Ca {\sc ii} 3933 and He {\sc i} 4026 have similar equivalent widths,
indicating that the summed HVS spectrum has a spectral type of B8 - B9.  This late B 
spectral type is consistent with our color selection and our previous spectral
classifications.

	Amusingly, we also see two Ca {\sc ii} K lines in Figure \ref{fig:spectrum}.  
One Ca {\sc ii} K line is from the atmospheres of the HVSs at 3933 \AA, and the
other is from the local interstellar medium in the direction of HVS8 and HVS9.  The
interstellar Ca {\sc ii} K line appears shifted to 3926 \AA\ in the rest-frame of
HVS8 and HVS9.  The difference between the two Ca {\sc ii} K lines, $\Delta \lambda
/ \lambda \sim$ 550 km s$^{-1}$, visibly indicates the large space motion of the
HVSs.

\section{ NATURE OF THE HYPERVELOCITY STARS }

	We use the observed velocity distribution (Figure \ref{fig:hist}) to show
that HVSs must be short-lived.  Using stellar evolution arguments and orbit
calculations, we argue that the HVSs are likely main sequence stars.

\subsection{ The Asymmetric Velocity Distribution }

	The HVSs are moving away from us with large positive radial velocities
$400<v_{\rm rf}<700$ km s$^{-1}$.  As stated earlier, the escape velocity of the Galaxy
at 50 kpc is $\sim$300 km s$^{-1}$, or 360 km s$^{-1}$ in the \citet{bromley06}
model (see Figure \ref{fig:ben}).  Therefore the absence of objects with
$v_{\rm rf}<-400$ km s$^{-1}$ is no surprise:  unbound HVSs will never come back.

	The significant excess of $+300$ km s$^{-1}$ stars compared to $-300$ km
s$^{-1}$ stars shows that the possibly bound HVSs {\it must be short-lived stars:}
every star traveling at $+300$ km s$^{-1}$ must eventually fall back onto the Galaxy
at roughly $-300$ km s$^{-1}$.  \citet{kollmeier07} and \citet{yu07} also make this
point.  The return time depends on the apocenter of the orbits, and ranges from 800
Myr to 2400 Myr for the range of distances and velocities we observe.  Although a
HVS's radial orbit may be altered by the asphericity of the Galactic potential
\citep{yu07}, the effect on radial velocity should be minor at the distances probed
by our survey.

\subsection{ HVSs are Main Sequence Stars }

	The stars we observe have the spectral types of B stars, but they may be
main sequence or post-main sequence stars.
	After a main sequence star exhausts its hydrogen fuel, it evolves up the red
giant branch, ignites helium, and becomes a blue horizontal branch (BHB) star.  All
stars with masses capable of igniting helium burning, including turn-off stars in
the halo, spend of order $10^8$ yr as BHB stars \citep{yi97}.
	The ambiguity in identifying BHB stars arises because the horizontal branch
overlaps the main sequence at the effective temperatures of our late B-type stars,
$T_{\rm eff}\sim10^4$ K.  Thus we cannot discriminate BHB and main sequence stars by
surface gravity alone.

	In Paper I, we use kinematics and metallicities to show that most of the
B-type stars are consistent with a halo population of post-main sequence stars or
blue stragglers in the halo.  The HVSs are a different class of objects.  Thus the
question stands:  are the HVSs 3 - 4 M$_{\sun}$ main sequence stars or BHB stars?

	The colors we observe preferentially sample hot BHB stars with small
hydrogen envelopes.  Hot BHB stars are present in only 25\% of globular clusters
\citep{lee07}.  One may wonder whether hot BHB stars are common in a Galactic center
population with solar abundance \citep{carr00, ramirez00, cunha07}.  
Theoretical evolutionary tracks allow for BHB stars at all metallicities
\citep{yi97}.  Hot BHB stars may, in fact, be more common at high metallicities
because increased opacity may cause increased mass loss in the red giant branch
phase \citep{faulkner72}.  Evidence for this picture comes from the metal-rich
[Fe/H]$\simeq$+0.4 open cluster NGC 6791, where a third of its helium-burning stars
are hot BHB stars \citep{yong00}.

	It is unlikely, however, that a HVS is ejected during the horizontal branch
phase.  In the red giant phase, prior to becoming a BHB star, a star swells to a
radius of $\sim$1 AU.  HVS ejections from a single MBH require stellar binaries with
separations of 0.1 - 1 AU \citep{hills88, bromley06}.  Thus any progenitor HVS
system experiences a common envelope phase that in-spirals the binary system over
$\lesssim10^3$ yrs \citep{webbink07} that probably prevents a HVS ejection by the
time the star reaches the BHB phase.  If HVSs are BHB stars, they must be ejected as
main sequence stars.  This condition is relaxed for single stars ejected by a pair
of MBHs or an in-spiraling IMBH.

\begin{figure}          
 \includegraphics[width=3.25in]{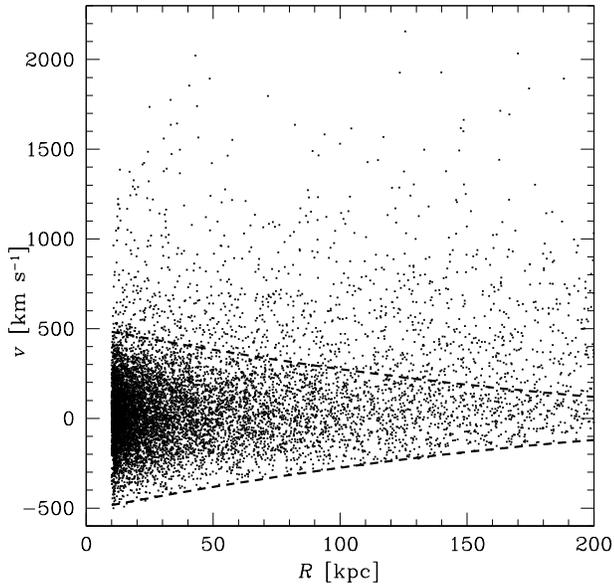}   
 \caption{ \label{fig:ben}
        Distribution of distance and velocity of simulated 1 M$_\sun$ HVSs from the
\citet{bromley06} ejection model.  The dashed line indicates the approximate escape
velocity of the potential.  In the range $10<R<50$ kpc, 95\% of the simulated 1
M$_{\sun}$ HVSs are bound, and 2/3 have crossed the Galactic center multiple times.  
If HVSs are BHB stars, we should thus find approximately equal numbers of stars at
$+300$ km s$^{-1}$ and $-300$ km s$^{-1}$, contrary to the observations.}
 \end{figure}

	However, the lifetime constraint derived from the observed velocities means
that the HVSs cannot be BHB stars descended from stars of mass less than 2 M$_{\sun}$.  
The progenitors of BHB stars span a wide range of stellar masses and include stars
with main sequence lifetimes up to a Hubble time.  Thus low-mass HVSs ejected many
Gyr ago may be observed as BHB stars in our survey.  In Figure \ref{fig:ben} we
illustrate the spatial and velocity distribution of such a population from a
simulation of 1 M$_{\sun}$ HVS ejections.  \citet{bromley06} describes the details
of the ejection model.  In brief, we disrupt equal mass binaries using the Hill's
mechanism, and assume that the 1 M$_{\sun}$ stars are ejected at a random time
during their main sequence lifetime.
	For reference, we also show the approximate escape velocity of the potential
model with dashed-lines (Figure \ref{fig:ben}).

	Our magnitude-limited survey detects BHB stars in the range $10<R<50$ kpc.  
Over this range of $R$, 95\% of the simulated 1 M$_{\sun}$ HVSs are bound; 2/3 of
them have crossed the Galactic center multiple times.
	Thus, if HVSs are BHB stars, we should observe approximately equal numbers
of BHB stars at $+300$ km s$^{-1}$ and $-300$ km s$^{-1}$, contrary to the
observations (Figure \ref{fig:hist}).  We simply don't see BHB stars from 
long-lived progenitors falling back onto the Galaxy with large negative radial 
velocities.

	Thus HVSs must have main sequence lifetimes less than the $\sim$1 Gyr
turn-around time.  Given the color-selection of our survey, the HVSs are either 3 -
4 M$_{\sun}$ main sequence stars, or possibly hot BHB stars evolved from $M>2$
M$_{\sun}$ progenitors.
	Our understanding of $M>2$ M$_{\sun}$ BHB stars is limited by the absence of
stellar evolution tracks that connect the zero age main sequence mass to the
horizontal branch core mass.
	Main sequence stars, on the other hand, match the current understanding of
the Galactic center:  \citet{maness07} find evidence for a top-heavy initial mass
function in the Galactic center, and \citet{eisenhauer05} show that the stars
presently in orbit around the central MBH are main sequence B stars, some of which
may be the former companions of the HVSs \citep{ginsburg06, ginsburg07}.
	The 3 - 4 M$_{\sun}$ main sequence stars sampled by our survey have
lifetimes of 160 - 350 Myr, consistent with the absence of bound HVSs falling back
onto the Galaxy with large negative velocities. 
	We conclude HVSs are likely main sequence stars.

\section{ SPACE DENSITY OF HYPERVELOCITY STARS }

	Our essentially complete survey places interesting constraints on the space 
density of HVSs.  
	The presence of a MBH in the Galactic center inevitably ejects a fountain of
HVSs from the Galaxy.  If HVS ejections are continuous and isotropic, their space
density should be proportional to $R^{-2}$ \citep[see also][]{bromley07}.  The
volume sampled by our survey is proportional to $R^3$.  Thus, in the simplest
picture, we expect the number of HVSs in our survey to have the dependence
$N(<R)\propto R$.  The actual space density depends on the luminosities of the HVSs.  
We use the space density to predict HVS discoveries for future surveys.

\begin{figure}          
 \includegraphics[width=3.25in]{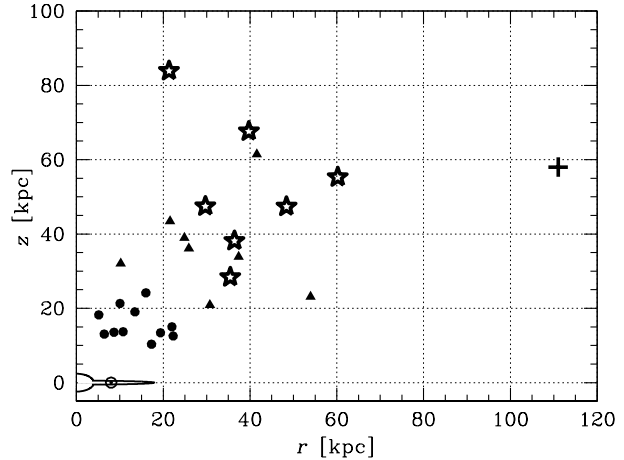}
 \caption{ \label{fig:gal}
        Location of our 7 HVSs ({\it stars}), the possibly bound HVSs ({\it circles,
triangles}), and HVS1 ({\it plus sign}) assuming they are main sequence B stars.
        $z$ is the distance above the Galactic plane and $r$ is the distance along
the Galactic plane, such that $R=(r^2 + z^2)^{-0.5}$.
        For reference, we sketch the Milky Way and the Sun at $(r,z)=(8,0)$ kpc. }
 \end{figure}

\subsection{ HVS Luminosities and Distances }

	Having established that HVSs are likely main sequence stars, we estimate
their intrinsic luminosities from \citet{schaller92} stellar evolutionary tracks for
solar metallicity stars.
	Five of the HVSs in our survey have colors (see Figure \ref{fig:ugr}) and
spectra consistent with 3 M$_{\sun}$ stars; the two bluest HVSs are consistent with
4 M$_{\sun}$ stars.  Using bolometric corrections from \citet{kenyon95}, we estimate
$M_V$(3 M$_{\sun})\simeq-0.3$ and $M_V$(4 M$_{\sun})\simeq-0.9$.
	These absolute magnitudes put the HVSs at distances in the range $45 < R <
90$ kpc.

\begin{figure}          
 \includegraphics[width=3.25in]{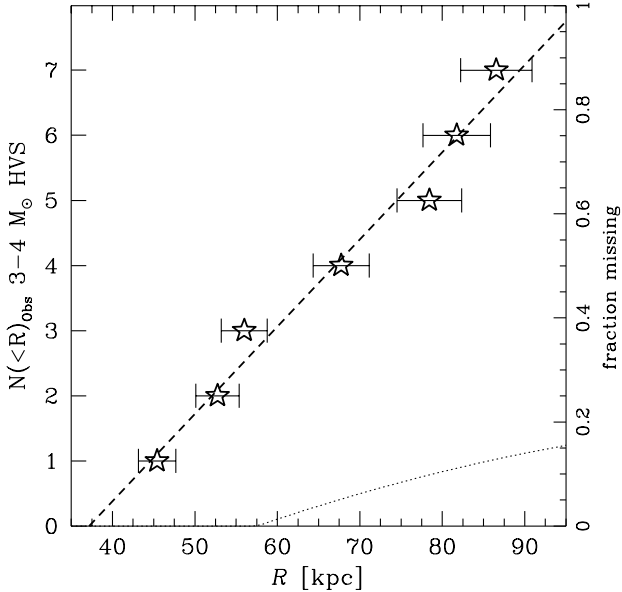}
 \caption{ \label{fig:densa}
        Observed distribution of HVSs ({\it stars}) vs.\ Galactocentric radius $R$.
Errorbars indicate 10\% luminosity uncertainties.  Dotted line shows the fraction of
HVSs with travel times exceeding their lifetime at distance $R$, indicated on the
right-hand vertical axis.  The best-fit line through the data ({\it dashed line}),  
corrected for incompleteness, has slope $0.159\pm0.017$ kpc$^{-1}$. }
 \end{figure}

	Figure \ref{fig:gal} illustrates the HVS's location in the outer halo of the
Galaxy.  The possibly bound HVSs, in contrast, are mostly found at closer distances;  
there are 11 possibly bound HVSs and no unbound HVS in the bright FAST sample. Given
a $N(<R)\propto R$ distribution, one may wonder why there are no unbound HVSs
nearby.  Projection effects may provide one explanation.  Consider SDSS
J144955.58+310351.4 (highlighted in Paper II), a possibly bound HVS located at
$R\simeq17$ kpc and traveling $v_{\rm rf}=+447$ km s$^{-1}$.  If SDSS
J144955.58+310351.4 has a purely radial trajectory, its true space velocity is +525
km s$^{-1}$ and it may thus be unbound.  But because we do not know which of the
possibly bound HVSs are true HVSs, we use only the well-defined sample of unbound
HVSs below.

	We next compare the HVS distances with the range of distances sampled by our
magnitude-limited survey.  Our survey is complete over the magnitude range
$15<g'_0<19.5$, corresponding to heliocentric distances of $12<d<100$ kpc and
$16<d<130$ kpc for 3 M$_{\sun}$ and 4 M$_{\sun}$ stars, respectively.  The range of
Galactocentric $R$ varies with Galactic latitude and longitude on the sky.  Over our
7300 deg$^2$ survey region, we are complete over the range $20<R<93$ kpc and
$24<R<124$ kpc for 3 M$_{\sun}$ and 4 M$_{\sun}$ stars, respectively.  The inner
edge of the MMT survey, in which the HVSs are found, is located at $R=38$ kpc and
$R=48$ kpc for 3 M$_{\sun}$ and 4 M$_{\sun}$ stars, respectively.

	Figure \ref{fig:densa} plots the cumulative distribution of distances for
the 7 HVSs in this survey.  The distribution is remarkably consistent with a
continuous ejection process:  a KS test finds a 0.96 likelihood of drawing the
distribution of HVSs from a linear distribution $N(<R)\propto R$.  This conclusion
is independent of the absolute scale of $R$.

\subsection{ Lifetime Correction }

	Before computing the space density of HVSs from Figure \ref{fig:densa}, we
must correct for the number of HVSs missing in our survey because of their finite
lifetimes.  Using the ejection models of \citet{bromley06}, we calculate the
fraction of stars with $v > v_{esc}$ at a given $R$ with travel times exceeding
their lifetimes.  The dotted line in Figure \ref{fig:densa} shows the weighted
average of the 3 and 4 M$_{\sun}$ correction factors, appropriate for our set of
HVSs.

	The lifetime correction is negligible for 3 M$_{\sun}$ HVSs in our magnitude
range, but it is substantial for the faintest 4 M$_{\sun}$ HVSs.  The two 4
M$_{\sun}$ HVSs in our survey are located at $R=56$ kpc and $R=82$ kpc; given a
$N(<R)\propto R$ dependence, we expect to find one or two more 4 M$_{\sun}$ HVSs out
to $R=124$ kpc.  However, the \citet{bromley06} ejection models show that 50\% of
unbound 4 M$_{\sun}$ HVSs are dead by $R=90$ kpc, and 67\% are dead by $R=110$ kpc.  
Thus the absence of 4 M$_{\sun}$ HVSs with $R>90$ kpc is consistent with the
interpretation that they are main sequence stars with short, 160 Myr lifetimes.  
The 3 M$_{\sun}$ HVSs, on the other hand, fill the region where we can detect them,
consistent with their longer, 350 Myr main sequence lifetimes.

\begin{figure}          
 \includegraphics[width=3.25in]{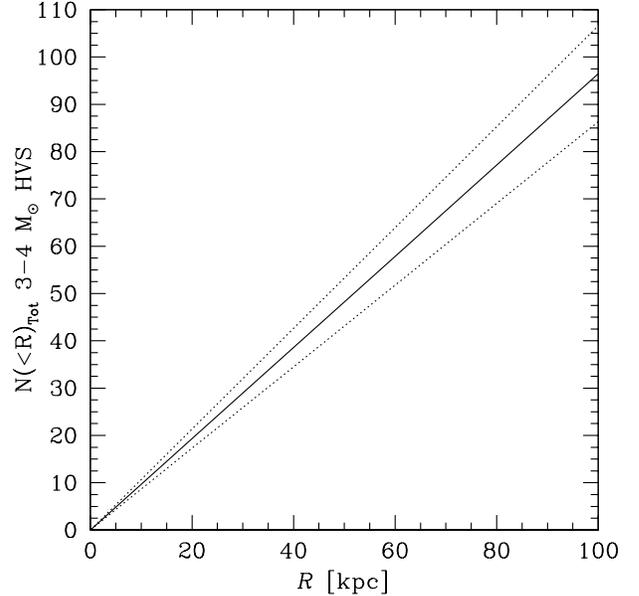}
 \caption{ \label{fig:densb}
        Cumulative number of 3 - 4 M$_{\sun}$ HVSs vs.\ Galactocentric radius $R$. 
Over $4\pi$ str, the space density of 3 - 4 M$_{\sun}$ HVSs is $\rho(R) =
(0.077\pm0.008) R^{-2}$ kpc$^{-3}$.  We infer there are $96\pm10$ 3 - 4 M$_{\sun}$
HVSs within $R<100$ kpc.}
 \end{figure}

\subsection{ Density Distribution }

	We derive the HVS space density from a least-squares fit to the
observations, corrected for lifetime.  The slope is $0.159\pm0.017$ kpc$^{-1}$ (see
Figure \ref{fig:densa}).  We estimate the uncertainty by assigning random
uncertainties to the HVSs, drawn from a Gaussian distribution with $\sigma=$10\% in
luminosity, and by re-fitting the line 10$^4$ times.  Because all 7 HVSs have
$g'_0>17$, they are drawn from the MMT survey and cover an effective area of 6800
deg$^2$ of sky (93\% of 7300 deg$^2$).  Corrected to $4\pi$ steradians, the space
density of 3 - 4 M$_{\sun}$ HVSs is thus $\rho(R) = (0.077\pm0.008) R^{-2}$
kpc$^{-3}$.

	Figure \ref{fig:densb} plots the cumulative number of 3 - 4 M$_{\sun}$ HVSs
in the Galaxy.  Our observations imply there are $96\pm10$ late B-type HVSs in the
sphere $R<100$ kpc, about 1 HVS per kpc. 

\subsection{ Prediction for Future Surveys }

	We can use the space density of 3 - 4 M$_{\sun}$ HVSs to predict HVS
detections in future surveys.  Our predictions make the following assumptions. 
	(1) We assume that the distribution of HVS velocities is the same for all
stellar masses.  Ejection velocity depends weakly on the mass of the stellar binary
$(m_1 + m_2)^{1/3}$ in the Hill's mechanism, but there is no such dependence for
stars ejected from a binary MBH.
	(2) We use the Salpeter mass function, $M^{-2.35}$ \citep{salpeter55}, and
the present-day mass function of the bulge, $M^{-4.5}$ for $M>1$ M$_{\sun}$ and
$M^{-2.35}$ for $M<1$ M$_{\sun}$ \citep{mezger99}, as two representative mass
functions for calculating the number ratios of stars of different stellar masses.
	(3) If HVSs are ejected by the Hill's mechanism, we implicitly assume
that the same fraction of stars are in compact binaries at all masses; no such
assumption is required for the binary MBH mechanism.
	(4) We use the \citet{girardi04} stellar isochrones of solar abundance to
obtain luminosities for stars at a given mass, or color.  We use the luminosities to
calculate survey volumes for a set of magnitude limits.
	(5) Finally, we assume observation of 100\% of stars in a given area of sky.

	\citet{kollmeier07} propose observing faint $19.5 < g' < 21.5$ stars near
the main-sequence turn-off $0.3 < (g'-i') < 1.1$ as the optimal strategy for finding
low-mass HVSs.  The $(g'-i')$ colors select $0.8 < M < 1.3$ M$_{\sun}$ solar
metallicity stars with absolute magnitudes $3.7 < M_V < 6.3$.  Using the present-day
mass function of the bulge, we estimate that there is 1 HVS per $\sim$50 deg$^2$ in
the proposed magnitude range, in excellent agreement with the 1 HVS per 45 deg$^2$
estimated by \citet{kollmeier07}.  However, a Salpeter mass function predicts an
order-of-magnitude lower density, 1 HVS per $\sim$500 deg$^2$.  The ratio of high-
to low-mass HVSs thus provides a sensitive measure of the stellar mass function near
the central MBH \citep{kollmeier07, bromley07}.

	SEGUE is an on-going survey that includes spectroscopy of 1144 stars along
$\sim$200 sightlines with the SDSS telescope \citep{adelman07b}.  Each sightline
covers 7 deg$^2$ of sky, for a total of 1400 deg$^2$ of spectroscopic coverage.  
Notably, SEGUE acquires spectra for 150 BHB/A-type stars per sightline over the
magnitude range $14<g'<20.5$.  The BHB/A color cut \citep{adelman07b} selects for
$1.5 < M < 2.5$ M$_{\sun}$ stars with absolute magnitudes $0.7 < M_V < 2.8$.  We
estimate SEGUE should find $\sim$5 to $\sim$15 HVSs for the Salpeter and present-day
mass functions, respectively.  However, if only a fraction of BHB/A stars are
observed over the 1400 deg$^2$ area, or if the spectra are not reliable to a depth
$g'=20.5$, our predictions may be over-estimates.

	SEGUE also targets F/G-type stars over the magnitude range $14<g'<20.2$.  
If we ignore low-metallicity targets as unlikely HVS candidates, SEGUE targets 50
F/G main sequence stars per sightline.  The color cut $0.2 <(g'-r')<0.48$
\citep{adelman07b} selects for $1.1 < M < 1.4$ M$_{\sun}$ stars with absolute
magnitudes $3.0 < M_V < 4.6$.  We estimate that 50 stars per sightline account for
$\sim$10\% of the total population of F/G stars, based on number counts of stars in
stripe 82.  We estimate that SEGUE should find $\sim$0.1 to $\sim$1 F/G-type HVS for
the Salpeter and present-day mass functions, respectively.  The F/G-type HVS
discovery rate is low because SEGUE is not complete for F/G stars and because F/G
stars have intrinsically low luminosity.  

	Regardless of stellar type, future HVS discoveries require surveys that 1)
sample large areas of sky $\gg10^2$ deg$^2$ and 2) sample large depths $R\gg10$ kpc.  
As a practical matter, the HVSs must also have a high contrast with respect to the
indigenous stellar population for efficient detection.  Our HVS survey is successful
because late B-type stars are luminous, and because the contamination from evolved
BHB stars and their low mass progenitors is minimal, $\sim$100 halo stars per HVS.

\begin{figure}          
 \includegraphics[width=3.25in]{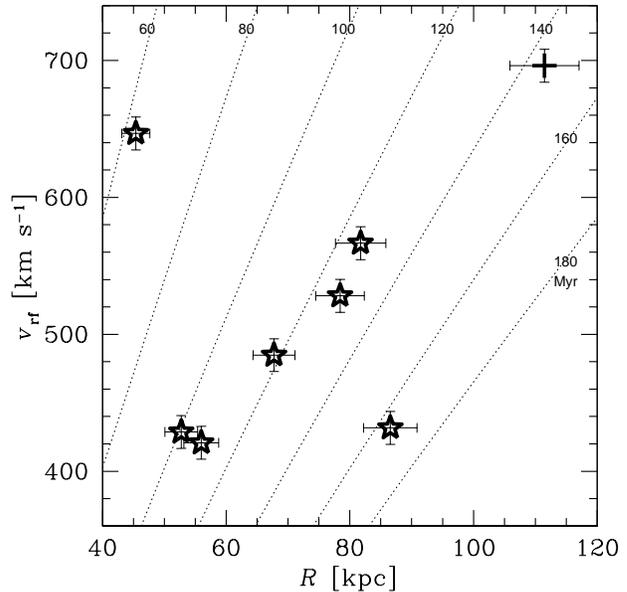}
 \caption{ \label{fig:travel}
        Minimum rest-frame velocity and distance of the HVSs ({\it stars}) and HVS1
({\it plus sign}).  Errorbars indicate the precision of the measurements; systematic
errors may be larger.  Travel times from the Galactic center ({\it dotted lines})
are calculated using the potential model of \citet{bromley06} and assuming the
minimum rest frame velocity $v_{rf}$ is the full space motion of the stars.  The
distribution of travel times statistically favor a constant ejection rate, though a
120 Myr-old burst of HVSs is also allowed.}
 \end{figure}

\section{ EJECTION HISTORY OF HYPERVELOCITY STARS }

	Knowing the distances and velocities of the HVSs, we can estimate their
travel times from the Galactic center.  Although we do not know the HVSs' true space
motions, at large distances their motion should be mostly radial.  Thus we calculate
travel times assuming the HVSs' minimum rest-frame velocities are their full space
motions.  This assumption means the travel time estimates are in fact upper limits.  
We use the potential model of \citet{bromley06} to correct for the deceleration of
the HVSs.  The deceleration correction reduces travel times by 10\% compared to
travel times calculated assuming constant velocity (as reported in Paper I).
	Table \ref{tab:hvs} lists travel time estimates for HVS4 - HVS10 and HVS1,
though we do not include HVS1 in this analysis. Figure \ref{fig:travel} plots the
distribution of travel times.

	The HVSs sample the ejection history of stars from the Galactic center over
the past $\sim$165 Myr.  The HVS travel times might favor a single burst, like that
expected from the in-spiral of a binary MBH, or a continuous ejection process, as
expected from Hill's mechanism.  Interestingly, 5 of the 7 HVS travel times appear
clustered around 120 Myr.  A Kolmogorov-Smirnov test finds a 0.8 likelihood of
drawing all 7 HVSs from a linear distribution in time, and a 0.2 likelihood of
drawing them from a single burst (e.g.\ travel times that fall along the dotted line
marked 120 Myr in Figure \ref{fig:travel}).

	The statistics appear to favor a continuous ejection process.  Thus Hill's
mechanism can plausibly account for the entire set of observed HVSs.  A single burst
event is also allowed, however.  If HVSs were ejected from the in-spiral of a binary
MBH, the observations suggest such an event may have occurred $\sim$120 Myr ago.  
We caution that small number statistics prevent any strong conclusion; additional
HVS discoveries are required to properly constrain the ejection history of stars
from the Galactic center.

\section{CONCLUSIONS}

	Our targeted HVS survey, a spectroscopic survey of stars with B-type colors,
is 96.5\% complete over 7300 deg$^2$ of the SDSS DR5.  We report 3 new HVS
discoveries, for a total of 7 HVSs with $15<g'_0<19.5$ over a sixth of the sky.  In
addition, we find 19 possibly bound HVSs with velocities $275<v_{\rm rf}<400$ km
s$^{-1}$.

	The significant excess of $+300$ km s$^{-1}$ stars compared to $-300$ km
s$^{-1}$ stars shows that the HVSs must be short-lived.  Any population of post-main
sequence HVSs must contain stars falling back onto the Galaxy, contrary to the
observations.  We conclude the HVSs are likely 3 - 4 M$_{\sun}$ main sequence stars,
although hot BHB stars from $M>2$ M$_{\sun}$ progenitors are also a possibility.
	Observations of stars in the central 10 pc of the Galaxy are now probing
$\sim$3 M$_{\sun}$ main sequence stars, some of which may feed the MBH and produce
HVSs.  Thus there is a need to better understand the BHB phase of 2 - 3 M$_{\sun}$
stars in the context of HVSs and Galactic center research.

	The spatial distribution of HVSs supports the main sequence interpretation.  
The longer-lived 3 M$_{\sun}$ HVSs fill our survey volume; the shorter-lived 4
M$_{\sun}$ HVSs are missing at faint magnitudes.  The spatial distribution of HVSs
is remarkably consistent with a $N(<R)\propto R$ distribution, as expected for a
continuous ejection process.  The density of 3 - 4 M$_{\sun}$ HVSs is $\rho(R) =
(0.077\pm0.008) R^{-2}$ kpc$^{-3}$, implying a total of $96\pm10$ such HVSs within
$R<100$ kpc.

	A sample of $\gtrsim$ 100 HVSs is required to discriminate among HVS
ejection mechanisms and Galactic potential models \citep{sesana07}.  Thus enough 3 -
4 M$_{\sun}$ HVSs may exist in the Galaxy to test the ejection and potential models,
especially if the bound HVSs can be used.  SEGUE and other future surveys will find
more HVSs and will allow further use of this important class of objects to connect
the Galactic center to the Galactic halo.

\acknowledgements

	We thank M.\ Alegria, J.\ McAfee, and A.\ Milone for their assistance with
observations obtained at the MMT Observatory, a joint facility of the Smithsonian
Institution and the University of Arizona.  We thank the referee for comments that
improved this paper.  This project makes use of data products from the Sloan Digital
Sky Survey, which is managed by the Astrophysical Research Consortium for the
Participating Institutions.  This research has made use of NASA's Astrophysics Data
System Bibliographic Services.  This work was supported by the Smithsonian
Institution.

{\it Facilities:} {MMT (Blue Channel Spectrograph)} {FLWO:1.5m 
(FAST Spectrograph)}


\begin{thebibliography}{76}
\expandafter\ifx\csname natexlab\endcsname\relax\def\natexlab#1{#1}\fi

\bibitem[{{Adelman-McCarthy} {et~al.}(2007{\natexlab{a}})}]{adelman07}
{Adelman-McCarthy}, J.~K. {et~al.} 2007{\natexlab{a}}, \apjs, in press

\bibitem[{{Adelman-McCarthy} {et~al.}(2007{\natexlab{b}})}]{adelman07b}
---. 2007{\natexlab{b}}, preprint astro-ph/0707.3413

\bibitem[{{Aerts} {et~al.}(1999)}]{aerts99}
{Aerts}, C. {et~al.} 1999, \aap, 343, 872

\bibitem[{{Baumgardt} {et~al.}(2006){Baumgardt}, {Gualandris}, \& {Portegies
  Zwart}}]{baumgardt06}
{Baumgardt}, H., {Gualandris}, A., \& {Portegies Zwart}, S. 2006, \mnras, 372,
  174

\bibitem[{{Blaauw}(1961)}]{blaauw61}
{Blaauw}, A. 1961, \bain, 15, 265

\bibitem[{{Bromley} {et~al.}(2006){Bromley}, {Kenyon}, {Geller}, {Barcikowski},
  {Brown}, \& {Kurtz}}]{bromley06}
{Bromley}, B.~C., {Kenyon}, S.~J., {Geller}, M.~J., {Barcikowski}, E., {Brown},
  W.~R., \& {Kurtz}, M.~J. 2006, \apj, 653, 1194

\bibitem[{{Bromley} {et~al.}(2007){Bromley}, {Kenyon}, {Geller}, {Barcikowski},
  {Brown}, \& {Kurtz}}]{bromley07}
---. 2007, ApJ, submitted

\bibitem[{{Brown} {et~al.}(2003){Brown}, {Allende Prieto}, {Beers}, {Wilhelm},
  {Geller}, {Kenyon}, \& {Kurtz}}]{brown03}
{Brown}, W.~R., {Allende Prieto}, C., {Beers}, T.~C., {Wilhelm}, R., {Geller},
  M.~J., {Kenyon}, S.~J., \& {Kurtz}, M.~J. 2003, \aj, 126, 1362

\bibitem[{{Brown} {et~al.}(2005){Brown}, {Geller}, {Kenyon}, \&
  {Kurtz}}]{brown05}
{Brown}, W.~R., {Geller}, M.~J., {Kenyon}, S.~J., \& {Kurtz}, M.~J. 2005,
  \apjl, 622, L33

\bibitem[{{Brown} {et~al.}(2006{\natexlab{a}}){Brown}, {Geller}, {Kenyon}, \&
  {Kurtz}}]{brown06}
---. 2006{\natexlab{a}}, \apjl, 640, L35

\bibitem[{{Brown} {et~al.}(2006{\natexlab{b}}){Brown}, {Geller}, {Kenyon}, \&
  {Kurtz}}]{brown06b}
---. 2006{\natexlab{b}}, \apj, 647, 303~(Paper I)

\bibitem[{{Brown} {et~al.}(2007{\natexlab{a}}){Brown}, {Geller}, {Kenyon}, \&
  {Kurtz}}]{brown07c}
---. 2007{\natexlab{a}}, \apj, 666, 231

\bibitem[{{Brown} {et~al.}(2007{\natexlab{b}}){Brown}, {Geller}, {Kenyon},
  {Kurtz}, \& {Bromley}}]{brown07a}
{Brown}, W.~R., {Geller}, M.~J., {Kenyon}, S.~J., {Kurtz}, M.~J., \& {Bromley},
  B.~C. 2007{\natexlab{b}}, \apj, 660, 311~(Paper II)

\bibitem[{{Brown} {et~al.}(2007{\natexlab{c}}){Brown}, {Kewley}, \&
  {Geller}}]{brown07d}
{Brown}, W.~R., {Kewley}, L.~J., \& {Geller}, M.~J. 2007{\natexlab{c}}, \aj,
  submitted

\bibitem[{{Carr} {et~al.}(2000){Carr}, {Sellgren}, \& {Balachandran}}]{carr00}
{Carr}, J.~S., {Sellgren}, K., \& {Balachandran}, S.~C. 2000, \apj, 530, 307

\bibitem[{{Clewley} {et~al.}(2004){Clewley}, {Warren}, {Hewett}, {Norris}, \&
  {Evans}}]{clewley04}
{Clewley}, L., {Warren}, S.~J., {Hewett}, P.~C., {Norris}, J.~E., \& {Evans},
  N.~W. 2004, \mnras, 352, 285

\bibitem[{{Cunha} {et~al.}(2007){Cunha}, {Sellgren}, {Smith}, {Ramirez},
  {Blum}, \& {Terndrup}}]{cunha07}
{Cunha}, K., {Sellgren}, K., {Smith}, V.~V., {Ramirez}, S.~V., {Blum}, R.~D.,
  \& {Terndrup}, D.~M. 2007, preprint astro-ph/0707.2610

\bibitem[{{Davies} {et~al.}(2002){Davies}, {King}, \& {Ritter}}]{davies02}
{Davies}, M.~B., {King}, A., \& {Ritter}, H. 2002, \mnras, 333, 463

\bibitem[{{Demarque} \& {Virani}(2007)}]{demarque07}
{Demarque}, P. \& {Virani}, S. 2007, \aap, 461, 651

\bibitem[{{Dolphin} {et~al.}(2003){Dolphin}, {Saha}, {Skillman}, {Dohm-Palmer},
  {Tolstoy}, {Cole}, {Gallagher}, {Hoessel}, \& {Mateo}}]{dolphin03}
{Dolphin}, A.~E., {Saha}, A., {Skillman}, E.~D., {Dohm-Palmer}, R.~C.,
  {Tolstoy}, E., {Cole}, A.~A., {Gallagher}, J.~S., {Hoessel}, J.~G., \&
  {Mateo}, M. 2003, \aj, 125, 1261

\bibitem[{{Edelmann} {et~al.}(2005){Edelmann}, {Napiwotzki}, {Heber},
  {Christlieb}, \& {Reimers}}]{edelmann05}
{Edelmann}, H., {Napiwotzki}, R., {Heber}, U., {Christlieb}, N., \& {Reimers},
  D. 2005, \apjl, 634, L181

\bibitem[{{Eisenhauer} {et~al.}(2005)}]{eisenhauer05}
{Eisenhauer}, F. {et~al.} 2005, \apj, 628, 246

\bibitem[{{Fabricant} {et~al.}(1998){Fabricant}, {Cheimets}, {Caldwell}, \&
  {Geary}}]{fabricant98}
{Fabricant}, D., {Cheimets}, P., {Caldwell}, N., \& {Geary}, J. 1998, \pasp,
  110, 79

\bibitem[{{Faulkner}(1972)}]{faulkner72}
{Faulkner}, J. 1972, \apj, 173, 401

\bibitem[{{Fisher} \& {Tully}(1975)}]{fisher75}
{Fisher}, J.~R. \& {Tully}, R.~B. 1975, \aap, 44, 151

\bibitem[{{Fuentes} {et~al.}(2006){Fuentes}, {Stanek}, {Gaudi}, {McLeod},
  {Bogdanov}, {Hartman}, {Hickox}, \& {Holman}}]{fuentes06}
{Fuentes}, C.~I., {Stanek}, K.~Z., {Gaudi}, B.~S., {McLeod}, B.~A., {Bogdanov},
  S., {Hartman}, J.~D., {Hickox}, R.~C., \& {Holman}, M.~J. 2006, \apjl, 636,
  L37

\bibitem[{{Ghez} {et~al.}(2005){Ghez}, {Salim}, {Hornstein}, {Tanner}, {Lu},
  {Morris}, {Becklin}, \& {Duchene}}]{ghez05}
{Ghez}, A.~M., {Salim}, S., {Hornstein}, S.~D., {Tanner}, A., {Lu}, J.~R.,
  {Morris}, M., {Becklin}, E.~E., \& {Duchene}, G. 2005, \apj, 620, 744

\bibitem[{{Ginsburg} \& {Loeb}(2006)}]{ginsburg06}
{Ginsburg}, I. \& {Loeb}, A. 2006, \mnras, 368, 221

\bibitem[{{Ginsburg} \& {Loeb}(2007)}]{ginsburg07}
---. 2007, \mnras, 376, 492

\bibitem[{{Girardi} {et~al.}(2002){Girardi}, {Bertelli}, {Bressan}, {Chiosi},
  {Groenewegen}, {Marigo}, {Salasnich}, \& {Weiss}}]{girardi02}
{Girardi}, L., {Bertelli}, G., {Bressan}, A., {Chiosi}, C., {Groenewegen},
  M.~A.~T., {Marigo}, P., {Salasnich}, B., \& {Weiss}, A. 2002, \aap, 391, 195

\bibitem[{{Girardi} {et~al.}(2004){Girardi}, {Grebel}, {Odenkirchen}, \&
  {Chiosi}}]{girardi04}
{Girardi}, L., {Grebel}, E.~K., {Odenkirchen}, M., \& {Chiosi}, C. 2004, \aap,
  422, 205

\bibitem[{{Gnedin} {et~al.}(2005){Gnedin}, {Gould}, {Miralda-Escud{\'e}}, \&
  {Zentner}}]{gnedin05}
{Gnedin}, O.~Y., {Gould}, A., {Miralda-Escud{\'e}}, J., \& {Zentner}, A.~R.
  2005, \apj, 634, 344

\bibitem[{{Gualandris} \& {Portegies Zwart}(2007)}]{gualandris07}
{Gualandris}, A. \& {Portegies Zwart}, S. 2007, \mnras, 376, L29

\bibitem[{{Gualandris} {et~al.}(2005){Gualandris}, {Portegies Zwart}, \&
  {Sipior}}]{gualandris05}
{Gualandris}, A., {Portegies Zwart}, S.~P., \& {Sipior}, M.~S. 2005, \mnras,
  363, 223

\bibitem[{{Hills}(1988)}]{hills88}
{Hills}, J.~G. 1988, \nat, 331, 687

\bibitem[{{Hirsch} {et~al.}(2005){Hirsch}, {Heber}, {O'Toole}, \&
  {Bresolin}}]{hirsch05}
{Hirsch}, H.~A., {Heber}, U., {O'Toole}, S.~J., \& {Bresolin}, F. 2005, \aap,
  444, L61

\bibitem[{{Kenyon} \& {Hartmann}(1995)}]{kenyon95}
{Kenyon}, S.~J. \& {Hartmann}, L. 1995, \apjs, 101, 117

\bibitem[{{Kewley} {et~al.}(2007){Kewley}, {Brown}, {Geller}, {Kenyon}, \&
  {Kurtz}}]{kewley07}
{Kewley}, L.~J., {Brown}, W.~R., {Geller}, M.~J., {Kenyon}, S.~J., \& {Kurtz},
  M.~J. 2007, \aj, 133, 882

\bibitem[{{Kilic} {et~al.}(2007{\natexlab{a}}){Kilic}, {Allende Prieto},
  {Brown}, \& {Koester}}]{kilic07}
{Kilic}, M., {Allende Prieto}, C., {Brown}, W.~R., \& {Koester}, D.
  2007{\natexlab{a}}, \apj, 660, 1451

\bibitem[{{Kilic} {et~al.}(2007{\natexlab{b}}){Kilic}, {Brown}, {Allende
  Prieto}, {Pinsonneault}, \& {Kenyon}}]{kilic07b}
{Kilic}, M., {Brown}, W.~R., {Allende Prieto}, C., {Pinsonneault}, M., \&
  {Kenyon}, S. 2007{\natexlab{b}}, \apj, 664, 1088

\bibitem[{{Kollmeier} \& {Gould}(2007)}]{kollmeier07}
{Kollmeier}, J.~A. \& {Gould}, A. 2007, \apj, 664, 343

\bibitem[{{Kurtz} \& {Mink}(1998)}]{kurtz98}
{Kurtz}, M.~J. \& {Mink}, D.~J. 1998, \pasp, 110, 934

\bibitem[{{Lee} {et~al.}(2007){Lee}, {Gim}, \& {Casetti-Dinescu}}]{lee07}
{Lee}, Y.-W., {Gim}, H.~B., \& {Casetti-Dinescu}, D.~I. 2007, \apjl, 661, L49

\bibitem[{{Leonard}(1991)}]{leonard91}
{Leonard}, P.~J.~T. 1991, \aj, 101, 562

\bibitem[{{Leonard}(1993)}]{leonard93}
{Leonard}, P.~J.~T. 1993, in ASP Conf.\ Ser.\ 45, Luminous High-Latitude Stars,
  ed. D.~Sasselov, 360

\bibitem[{{Levin}(2006)}]{levin06}
{Levin}, Y. 2006, \apj, 653, 1203

\bibitem[{{Lu} {et~al.}(2007){Lu}, {Yu}, \& {Lin}}]{lu07}
{Lu}, Y., {Yu}, Q., \& {Lin}, D.~N.~C. 2007, \apjl, 666, L89

\bibitem[{{Maness} {et~al.}(2007)}]{maness07}
{Maness}, H. {et~al.} 2007, \apj, accepted

\bibitem[{{Mateo}(1998)}]{mateo98}
{Mateo}, M.~L. 1998, \araa, 36, 435

\bibitem[{{Mathias} {et~al.}(2001){Mathias}, {Aerts}, {Briquet}, {De Cat},
  {Cuypers}, {Van Winckel}, \& {Le Contel}}]{mathias01}
{Mathias}, P., {Aerts}, C., {Briquet}, M., {De Cat}, P., {Cuypers}, J., {Van
  Winckel}, H., \& {Le Contel}, J.~M. 2001, \aap, 379, 905

\bibitem[{{M{\'e}ndez} {et~al.}(2002){M{\'e}ndez}, {Davis}, {Moustakas},
  {Newman}, {Madore}, \& {Freedman}}]{mendez02}
{M{\'e}ndez}, B., {Davis}, M., {Moustakas}, J., {Newman}, J., {Madore}, B.~F.,
  \& {Freedman}, W.~L. 2002, \aj, 124, 213

\bibitem[{{Merritt}(2006)}]{merritt06}
{Merritt}, D. 2006, \apj, 648, 976

\bibitem[{{Mezger} {et~al.}(1999){Mezger}, {Zylka}, {Philipp}, \&
  {Launhardt}}]{mezger99}
{Mezger}, P.~G., {Zylka}, R., {Philipp}, S., \& {Launhardt}, R. 1999, \aap,
  348, 457

\bibitem[{{O'Leary} \& {Loeb}(2007)}]{oleary07}
{O'Leary}, R.~M. \& {Loeb}, A. 2007, preprint astro-ph/0609046

\bibitem[{{Perets} \& {Alexander}(2007)}]{perets07b}
{Perets}, H.~B. \& {Alexander}, T. 2007, preprint astro-ph/0705.2123

\bibitem[{{Perets} {et~al.}(2007){Perets}, {Hopman}, \& {Alexander}}]{perets07}
{Perets}, H.~B., {Hopman}, C., \& {Alexander}, T. 2007, \apj, 656, 709

\bibitem[{{Piotto} {et~al.}(1994){Piotto}, {Capaccioli}, \&
  {Pellegrini}}]{piotto94}
{Piotto}, G., {Capaccioli}, M., \& {Pellegrini}, C. 1994, \aap, 287, 371

\bibitem[{{Portegies Zwart}(2000)}]{portegies00}
{Portegies Zwart}, S.~F. 2000, \apj, 544, 437

\bibitem[{{Poveda} {et~al.}(1967){Poveda}, {Ruiz}, \& {Allen}}]{poveda67}
{Poveda}, A., {Ruiz}, J., \& {Allen}, C. 1967, Bol.\ Obs\ Tonantzintla
  Tacubaya, 4, 860

\bibitem[{{Ram{\'{\i}}rez} {et~al.}(2000){Ram{\'{\i}}rez}, {Sellgren}, {Carr},
  {Balachandran}, {Blum}, {Terndrup}, \& {Steed}}]{ramirez00}
{Ram{\'{\i}}rez}, S.~V., {Sellgren}, K., {Carr}, J.~S., {Balachandran}, S.~C.,
  {Blum}, R., {Terndrup}, D.~M., \& {Steed}, A. 2000, \apj, 537, 205

\bibitem[{{Sakamoto} {et~al.}(2003){Sakamoto}, {Chiba}, \&
  {Beers}}]{sakamoto03}
{Sakamoto}, T., {Chiba}, M., \& {Beers}, T.~C. 2003, \aap, 397, 899

\bibitem[{{Salpeter}(1955)}]{salpeter55}
{Salpeter}, E.~E. 1955, \apj, 121, 161

\bibitem[{{Sch{\" o}del} {et~al.}(2003){Sch{\" o}del}, {Ott}, {Genzel},
  {Eckart}, {Mouawad}, \& {Alexander}}]{schodel03}
{Sch{\" o}del}, R., {Ott}, T., {Genzel}, R., {Eckart}, A., {Mouawad}, N., \&
  {Alexander}, T. 2003, \apj, 596, 1015

\bibitem[{{Schaller} {et~al.}(1992){Schaller}, {Schaerer}, {Meynet}, \&
  {Maeder}}]{schaller92}
{Schaller}, G., {Schaerer}, D., {Meynet}, G., \& {Maeder}, A. 1992, \aaps, 96,
  269

\bibitem[{{Sesana} {et~al.}(2006){Sesana}, {Haardt}, \& {Madau}}]{sesana06}
{Sesana}, A., {Haardt}, F., \& {Madau}, P. 2006, \apj, 651, 392

\bibitem[{{Sesana} {et~al.}(2007{\natexlab{a}}){Sesana}, {Haardt}, \&
  {Madau}}]{sesana07}
---. 2007{\natexlab{a}}, \apj, 660, 546

\bibitem[{{Sesana} {et~al.}(2007{\natexlab{b}}){Sesana}, {Haardt}, \&
  {Madau}}]{sesana07b}
---. 2007{\natexlab{b}}, \mnras, 379, L45

\bibitem[{{Tauris} \& {Takens}(1998)}]{tauris98}
{Tauris}, T.~M. \& {Takens}, R.~J. 1998, \aap, 330, 1047

\bibitem[{{Webbink}(2007)}]{webbink07}
{Webbink}, R.~F. 2007, preprint astro-ph/0704.0280

\bibitem[{{Wilkinson} \& {Evans}(1999)}]{wilkinson99}
{Wilkinson}, M.~I. \& {Evans}, N.~W. 1999, \mnras, 310, 645

\bibitem[{{Wolff}(1978)}]{wolff78}
{Wolff}, S.~C. 1978, \apj, 222, 556

\bibitem[{{Yi} {et~al.}(1997){Yi}, {Demarque}, \& {Kim}}]{yi97}
{Yi}, S., {Demarque}, P., \& {Kim}, Y.-C. 1997, \apj, 482, 677

\bibitem[{{Yong} {et~al.}(2000){Yong}, {Demarque}, \& {Yi}}]{yong00}
{Yong}, H., {Demarque}, P., \& {Yi}, S. 2000, \apj, 539, 928

\bibitem[{{Young}(2000)}]{young00}
{Young}, L.~M. 2000, \aj, 119, 188

\bibitem[{{Yu} \& {Madau}(2007)}]{yu07}
{Yu}, Q. \& {Madau}, P. 2007, \mnras, 379, 1293

\bibitem[{{Yu} \& {Tremaine}(2003)}]{yu03}
{Yu}, Q. \& {Tremaine}, S. 2003, \apj, 599, 1129

\end{thebibliography}


\appendix
\section{DATA TABLE}

	Table \ref{tab:obs} [available as tab2.tex in the source] lists the 381 new
observations from our HVS survey, excluding the extra-galactic objects.  206 HVS
candidates were observed with the MMT and 175 were observed with FAST (see \S 2).  
Table \ref{tab:obs} includes columns of RA and Dec coordinates (J2000), $g'$
apparent magnitude, $(u'-g')_0$ and $(g'-r')_0$ color, and our heliocentric velocity
$v_{\sun}$.  The column WD indicates whether the object is a white dwarf (WD=1) or a
B-type star (WD=0).

\begin{deluxetable}{cccccrr}		
\tabletypesize{\small}
\tablewidth{0pt}
\tablecaption{HVS SURVEY: NEW OBSERVATIONS\label{tab:obs}}
\tablecolumns{7}
\tablehead{
	\colhead{RA} & \colhead{Dec} &
	\colhead{$g'$} & \colhead{$(u'-g')_0$} & \colhead{$(g'-r')_0$} &
	\colhead{WD} & \colhead{$v_{\sun}$} \\
	\colhead{J2000} & \colhead{J2000} &
	\colhead{\small mag} & \colhead{\small mag} & \colhead{\small mag} &
	\colhead{} & \colhead{\small km s$^{-1}$}
}
	\startdata
 0:23:09.05 &  -0:33:42.0 & $16.277\pm0.020$ & $0.250\pm0.028$ & $-0.306\pm0.039$ & 1 & $ -13\pm31$ \\
 0:23:53.29 &  -1:04:46.4 & $18.304\pm0.016$ & $0.749\pm0.042$ & $-0.255\pm0.025$ & 0 & $  20\pm10$ \\
 0:43:50.55 &  -9:52:27.0 & $16.326\pm0.032$ & $0.832\pm0.038$ & $-0.358\pm0.038$ & 0 & $-117\pm11$ \\
	\enddata 
  \tablecomments{Table \ref{tab:obs} is presented in its entirety in the
electronic edition of the Astrophysical Journal.  A portion is shown here for
guidance and content.}
 \end{deluxetable}

\end{document}